\documentclass[a4paper,twocolumn]{emulateapj}
\usepackage{amstext}
\usepackage{apjfonts}
\usepackage{amsmath}
\usepackage{amssymb}
\usepackage{xcolor,xspace}
\usepackage{multirow,threeparttable,tabularx}
\usepackage[colorlinks=true,linkcolor=blue,citecolor=blue]{hyperref}

\newcommand{\brg}{Br$\gamma$\xspace}
\newcommand{\sa}{Sgr~A*\xspace}
\newcommand{\msun}{M_{\odot}\xspace}
\newcommand{\Sim}{\sim\!}
\newcommand{\tn}[1]{\tablenotemark{#1}}
\newcommand{\mr}[1]{\multirow{2}{*}{#1}}
\newcommand{\tabspace}{\rule{0pt}{1.0em}}

\begin{document}

\shorttitle{G2 from the Tidal Disruption of a Known Giant}

\shortauthors{Guillochon, Loeb, MacLeod, Ramirez-Ruiz}

\title{Possible Origin of the G2 Cloud from the Tidal Disruption of a Known Giant Star by Sgr A*}

\author{James Guillochon\altaffilmark{1,2}, Abraham Loeb\altaffilmark{1}, Morgan MacLeod\altaffilmark{3}, Enrico Ramirez-Ruiz\altaffilmark{3}}
\altaffiltext{1}{Harvard-Smithsonian Center for Astrophysics, The Institute for Theory and
Computation, 60 Garden Street, Cambridge, MA 02138, USA}
\altaffiltext{2}{Einstein Fellow}
\altaffiltext{3}{Department of Astronomy and
  Astrophysics, University of California, Santa Cruz, CA
  95064}

\email{jguillochon@cfa.harvard.edu}

\begin{abstract} 
The discovery of the gas cloud G2 on a near-radial orbit about \sa has prompted much speculation on its origin. In this {\it Letter}, we propose that G2 formed out of the debris stream produced by the removal of mass from the outer envelope of a nearby giant star. We perform hydrodynamical simulations of the returning tidal debris stream with cooling, and find that the stream condenses into clumps that fall periodically onto \sa. We propose that one of these clumps is the observed G2 cloud, with the rest of the stream being detectable at lower \brg emissivity along a trajectory that would trace from G2 to the star that was partially disrupted. By simultaneously fitting the orbits of S2, G2, and $\Sim$ 2,000 candidate stars, and by fixing the orbital plane of each candidate star to G2 (as is expected for a tidal disruption), we find that several stars have orbits that are compatible with the notion that one of them was tidally disrupted to produce G2. If one of these stars were indeed disrupted, it last encountered \sa hundreds of years ago, and has likely encountered \sa repeatedly. However, while these stars are compatible with the giant disruption scenario given their measured positions and proper motions, their radial velocities are currently unknown. If one of these stars' radial velocity is measured to be compatible with a disruptive orbit, it would strongly suggest its disruption produced G2.
\end{abstract}


\nocite{Turk:2011dd}

\section{Introduction}

Only 1\% of supermassive black holes accrete at rates comparable to their Eddington limits \citep{Ho:2008kz}. \sa, a $\Sim 4 \times 10^{6} \msun$ black hole lying at the center of our galaxy \citep{Ghez:1998id,Genzel:2000hr}, is no exception, and is thought to be accreting only $10^{-9}$ to $10^{-6} \msun \;{\rm yr}^{-1}$ \citep{Narayan:1998ef,Yuan:2014vq}. Given this small accretion rate, it was surprising to discover an Earth-mass cloud (G2) in a disruptive orbit about \sa \citep{Gillessen:2012kr}, as its destruction would deposit a mass around \sa at a rate comparable to its steady accretion rate.

Because G2's low density places it well within its tidal disruption radius $r_{\rm t}$ at the time of its discovery, it has been proposed that it requires a star at its center to continually replenish gas \citep{MurrayClay:2012fn}. However, the non-detection of a host star with ${\rm K} \lesssim 20$ limits the kinds of stars that can reside within it (\citealt{Phifer:2013bm,Meyer:2013tb}, although see \citealt{Eckart:2013tb}), and is comparable to the brightness expected for a young T-Tauri star, which could host a protoplanetary disk \citep{MurrayClay:2012fn} or produce a wind \citep{Ballone:2013ho,Scoville:2013iz}.

When a star is disrupted by a supermassive black hole, material that is removed from it collimates into a thin stream that feeds the black hole for centuries at a rate greater than \sa's steady accretion rate \citep{Guillochon:2013jj,Guillochon:2014in}. We propose that G2 is associated with a star that is not contained within it, but is rather a starless clump within a long stream resulting from the disruption of a star over two centuries ago. Because G2's periapse distance $r_{\rm p} \sim 100$ AU, this would necessarily be a large giant with $R_{\ast} \sim 1$ AU. In this scenario, G2 is one of dozens of clouds that have fallen onto \sa on similar orbits, with the accretion rate likely being larger in the past, closer to the time of disruption. Indeed, X-ray light echoes suggest that \sa was far more active centuries ago \citep{Ryu:2013vg}. We show that several giant stars \citep{Rafelski:2007gu,Schodel:2009jm,Yelda:2010ig,Do:2013fn}, some of which lie (in projection) along the path of a stream of clumps discovered in K that are spatially coincident with a \brg detection \citep{Gillessen:2013fb,Meyer:2013tb}, are both large enough to lose mass at G2's periapse, and possess a proper motion and position that are compatible with a tidally disruptive orbit.

The work presented in this {\it Letter} is composed of two parts: A demonstrative hydrodynamical simulation of the return of a debris stream resulting from a tidal disruption of a giant star, and a systematic search to find a star whose orbit is compatible with the giant disruption scenario. In Section \ref{sec:stream} we show that the debris stream resulting from a giant disruption can produce both a G2-like clump on a Keplerian trajectory, and an extended tail structure whose shape is determined by the range of binding energies within the tidal tail. In Section \ref{sec:fit} we present a systematic search for the associated giant star within our galactic center (GC). In Section \ref{sec:implications} we discuss implications of G2 being produced by the disruption of a giant.

\begin{figure*}
\centering\includegraphics[width=\linewidth,clip=true]{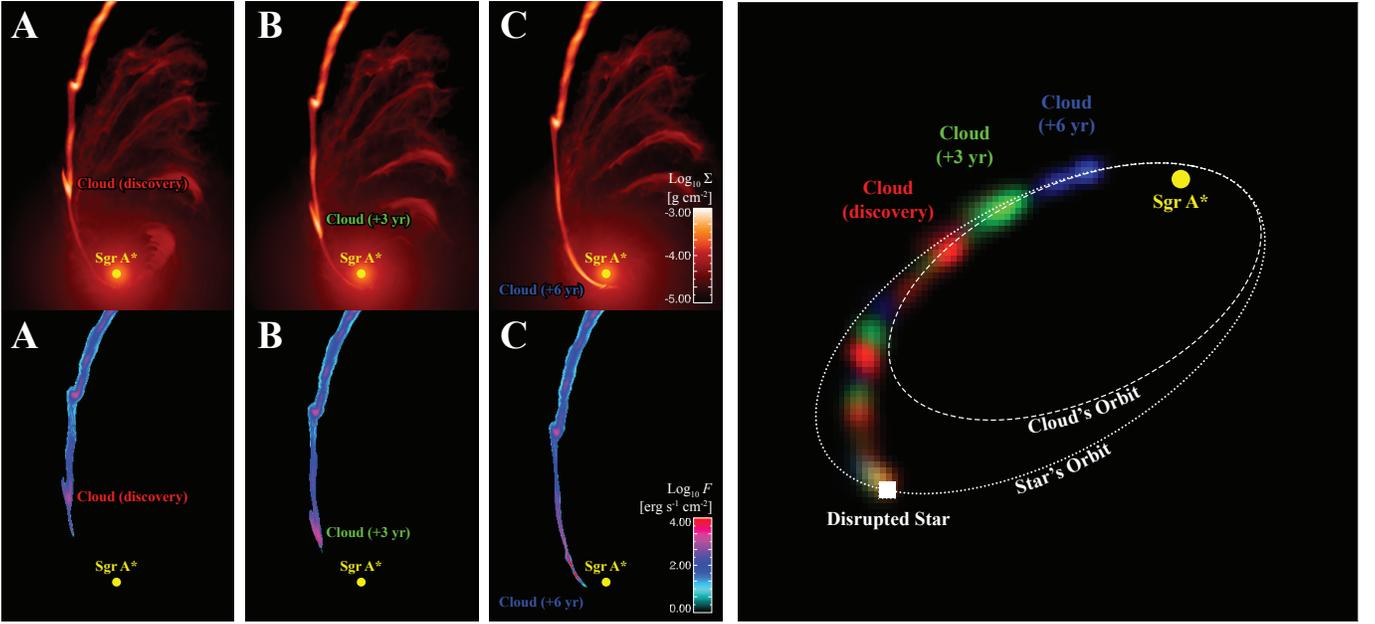}
\caption{Snapshots from a hydrodynamical simulation demonstrating the dynamics of a returning stream produced by the disruption of a giant star. In the left six panels (labeled A, B, and C) we show a time-sequence of the returning gas, with the column density $\Sigma$ shown in the top panels and total line-cooling flux $F$ shown in the bottom panels. Labeled in each panel is a prominent cloud that forms within the stream, falling onto the black hole over a period of $\sim 6$ yr. In the right panel we show $F_{{\rm Br}\gamma}$, with the simulation volume rotated such that its projection matches the best-fitting orientation for the G2 cloud as found by our MLA (Section \ref{sec:fit}), smoothed over 100 AU (33 AU per pixel). In addition to the lead cloud, there are a trail of clouds that follow it, but not on a path identical to leading cloud.}
\label{fig:stream}
\end{figure*}

\section{Simulation of a Tidal Stream in the GC}\label{sec:stream}
\subsection{Setup}
We perform a 3D hydrodynamical simulation of a debris stream returning to \sa in {\tt FLASH}\footnote{Movies available at \url{http://goo.gl/58iEFX}.}, a well-tested adaptive mesh code for astrophysical fluid problems \citep{Fryxell:2000em}. We presume a black hole of mass $M_{\rm h} = 4.3 \times 10^{6} \msun$ and set our background temperature profile based on measurements of diffuse X-ray emission in the GC \citep[Equation 2 of][]{Anninos:2012hr}. The GC is thought to be convectively unstable, which has been treated in previous simulations of G2 by either artificially reseting the background to a constant profile at every timestep \citep{Burkert:2012cy,Schartmann:2012cw,Ballone:2013ho}, or by simulating for a short period of time such that convective instability does not develop \citep{Abarca:2014ei}. Because we were concerned that the growth of instability within the stream might be affected by an ad-hoc relaxation scheme, our approach is different from the aforementioned works. As in \citet{Anninos:2012hr}, we set $\rho = 1.3 \times 10^{-21} \eta$ g cm$^{-3}$ (setting $\eta = 1$) at $r = 1.3 \times 10^{16}$ cm, but we presume a slightly steeper profile than the profile motivated by observations, $\rho \propto r^{-3/2}$. However, this configuration is stable to convection, obviating the need to artificially stabilize the background medium, and permitting unfettered evaluation of the growth of hydrodynamical stream instabilities. Our choice affects the distance from \sa at which instability will grow as growth is dependent on the ratio of densities between the two fluids, but should not affect the growth qualitatively.

When a star is partially disrupted, the mass it loses is distributed within a thin stream with a range of binding energies \citep{MacLeod:2013dh}. This stream remains thin as it leaves the vicinity of the black hole, so long as its evolution is adiabatic \citep{Kochanek:1994bn,Guillochon:2014in}. Unlike main sequence (MS) disruptions, giant disruptions produce streams that are initially much less dense, resulting in a stream that quickly becomes optically thin. Once optically thin, the stream's internal energy is set by the ionizing radiation from stars and gas in the surrounding GC environment, which floors the temperature of the gas component to $\Sim 10^{4}$ K. At the same time, the stream cools via recombination lines, with the cooling rate $\Lambda$ having a dependence on metallicity, temperature, and optical depth \citep{Sutherland:1993fr,Shcherbakov:2014bk}. We approximate $\Lambda$ as a Gaussian function of the temperature $T$ centered about $10^{5}$ K,
\begin{equation}
\Lambda = 10^{-21} \frac{\rho^{2}}{\mu_{\rm e}^{2} m_{\rm p}^{2}} \exp \left[-\frac{9}{2} \left(\log T - 5\right)^{2}\right]\;{\rm erg}\;{\rm cm}^{3}\;{\rm s}^{-1},\label{eq:cool}
\end{equation}
where $\rho$ is the density and $\mu_{\rm e}$ is the molecular weight per electron. In regions of sufficient density photoionization is balanced by recombination, and the stream equilibrates to $T \Sim 10^{4}$ K, which is used as a temperature floor in our simulation, and as the stream's initial temperature.

As the envelopes of giants that are removed upon disruption have negligible self-gravity \citep{MacLeod:2012cd}, the self-gravity of the stream is irrelevant, and thus the stream evolves isothermally in \sa's tidal gravity. The density profile of such a stream is
\begin{align}
\rho(s, r) &= \rho_{0}(r) \exp \left[-\left(\frac{s}{h}\right)^{2}\right]\label{eq:rho}\\
h &= \sqrt{\frac{2 r^3 k_{\rm b} T}{m_{\rm p} \mu_{\rm e} \mu_{\rm h}}},
\end{align}
where $s$ is the cylindrical distance from the center of the stream, $h$ is the cylindrical scale-height, $r$ is the distance to the black hole, $\mu_{\rm h} \equiv G M_{\rm h}$ is the standard gravitational parameter, and $\mu$ is the mean molecular weight. The density at the stream's core $\rho_{0}(r)$ is set by enforcing mass continuity through the cylinder assuming matter crosses through it at the Keplerian velocity $v_{\rm k}$, $\rho_{0}(r) = 2 \dot{M}/h^{2} v_{\rm k}$. We set $\dot{M} = 3 M_{\oplus}$ per decade, the rate implied by the accretion of one G2-sized cloud.

Our simulations feed matter to the black hole via a moving boundary condition with a cylindrical profile as determined by Equation \ref{eq:rho}, where the boundary lies at the apoapse of each fluid element, and is oriented perpendicular to the gas motion, set initially to the $-x$ direction. All fluid is initially placed on a Keplerian orbit with $r_{\rm p} = 200$ AU, but the boundary moves outwards such that its apoapse distance $r_{\rm a} = (t/t_{\rm t})^{2/3}$, where
\begin{equation}
t_{\rm t}~\equiv~2\pi \sqrt{r_{\rm p}^{3}/G M_{\rm h}}.
\end{equation}

\subsection{Results}
At some distance from \sa, the ambient gas pressure becomes competitive with the internal pressure of the stream, establishing pressure equilibrium between the stream and its surroundings. The internal pressure of the stream cannot equilibrate instantaneously, resulting in pulsational instabilities as the stream adjusts. At the same time, Kelvin-Helmholtz instabilities along the stream's boundaries in response to its free-fall relative to the stationary background \citep{Chandrasekhar:1961uk}. The combination of these two effects lead to over (under) densities developing at the crests (troughs) of the instabilities (Figure \ref{fig:stream}). Because the cooling rate is enhanced with greater density (Equation \ref{eq:cool}), the density perturbations in the stream are amplified, eventually resulting in a fragmented stream with dense clumps separated by tenths of arcseconds.

These clouds fall towards \sa in balance with the ambient pressure, and eventually succumb to tidal forces, stretching significantly (Figure \ref{fig:stream}, top three panels). Our simulations suggest that the total flux emitted by the clouds is approximately constant (Figure \ref{fig:stream}, bottom three panels), however this is likely a numerical effect resulting from our finite spatial resolution as clouds cannot collapse indefinitely, as they would in the pure isothermal case. In reality, the observed constant flux of G2 \citep{Gillessen:2012kr} is likely the result of collisional de-excitation as the volume density within the clouds increases above $\log p/k \sim 8$ \citep{Sutherland:1993fr}. This pressure threshold is less than the maximum seen within our simulations, suggesting the effect would halt isothermal collapse at a resolution comparable to that of our simulation.

As matter is continually fed to the black hole after a disruption, new clumps form continually form out of the stream. If G2 formed in this manner, it is one of many clumps that accreted onto \sa over the preceding centuries, and is unique only in the sense that it is the particular clump we happen to observe returning to \sa at the present epoch. Rather than each clump generating a strong bow shock in the ambient medium, the clumps follow a fixed path that is kept free of ambient gas by strands of lower-density (but higher-temperature) material between the clumps that also follow the same path (Figure \ref{fig:stream}, upper left panels). This would explain its non-detection in the radio \citep{Sadowski:2013ee,Akiyama:2013ty}. Each clump is tidally stretched and heated as it passes periapse, resulting in an increase in temperature above $10^{5}$ K, rendering them invisible in recombination lines, but potentially detectable in X-rays \citep{Anninos:2012hr}.

\section{Orbit fitting}\label{sec:fit}
If the formation of G2 is due to the partial disruption of a star, its initial orbital orientation, defined by the inclination $i$, argument of periapse $\omega$, longitude of the ascending node $\Omega$, and its initial pericenter distance $r_{\rm p}$ will be identical to the original star's. However, because the G2 clump originates from some piece of the debris stream that is more bound to \sa than the star that produced it, its specific orbital energy $\epsilon$ will be larger by a factor $\Delta \epsilon$, which means that the semimajor axis $a$ of G2 is restricted to values smaller than the star it originated from. Hence, if G2's orbital elements and the position/velocity of \sa were known, there is a single free parameter $a$.

As G2's orbital parameters and the position of \sa are only known to some precision, finding a star whose orbit is compatible with G2's orbit requires simultaneously determination of G2's orbital elements (six parameters), \sa's position, velocity, and mass (seven parameters), and the candidate star's $a$. Because the error bars on G2's position are too great to constrain \sa alone, we also simultaneously fit the orbital parameters of the short-period star S2 (six parameters), as is done in \citet{Phifer:2013bm} and \citet{Gillessen:2013bt}. This enables a precise determination of \sa's position, and allows one to calibrate the reference frame \citep{Gillessen:2009fg}. However, the errors in $M_{\rm h}$ and distance to the GC $R_{0}$ are rather large when using S2 alone, so we additionally use priors of $M_{\rm h} = 4.31 \pm 0.42 \times 10^{6} \msun$ and $R_{0} = 8.33 \pm 0.35$ kpc as determined by \citet{Gillessen:2009fn}. As in that paper, we assume a prior on \sa's radial velocity $v_{r} = 0 \pm 5$ km s$^{-1}$.

\begin{figure*}
\centering\includegraphics[width=\linewidth,clip=true]{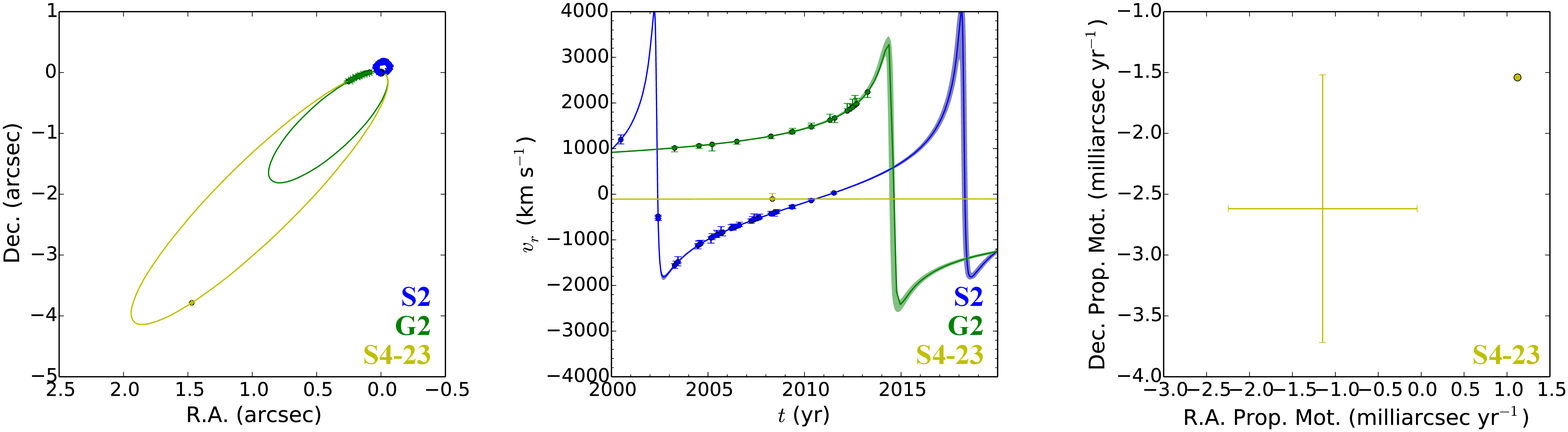}
\caption{Results from simultaneous fits of S2, G2, and S4-23, with the three objects being colored blue, green, and yellow respectively (\sa's position is shown with a black point). The points with error bars are the observational data, whereas the filled circles are the positions of the objects according to our model. The left panels show the sky-projected paths of the three objects for the maximum likelihood fit, with $x$ and $y$ corresponding to right ascension and declination. The middle panels show the velocity along the line of sight $v_{z}$, with the shaded regions indicating the 2-$\sigma$ model error bars in $v_{z}$ as a function of time (S4-23 $v_{z}$ provided by S. Gillessen, private communication). The right panels show the proper motion measurement of S4-23, with the error bars corresponding to observations and the filled circle to the model.\vspace{2.5em}}
\label{fig:orbits}
\end{figure*}

\begin{table*}
\scriptsize
\centering\begin{threeparttable}
\caption{Table of estimated orbital properties for G2 and candidate stars}
\begin{tabularx}{0.92\linewidth}{cccccccccc}
\hline\hline
Score\tn{a} & ID(s) & K & $P$ (yr) & $r_{\rm p}$ ($r_{\rm g}$) & $t_{\rm p, 0}$ (yr)\tn{b}
& $t_{\rm p, 1}$ (yr)\tn{c} & $a$ (mpc) & $e$ & $v_{z}$ (km s$^{-1}$)\tn{d}\\
\hline
\multicolumn{10}{c}{\citet{Yelda:2010ig}}\\
\hline
\mr{0} & G2 & --- & $518 \pm 41$ & $1600 \pm 97$ & $1496.79 \pm 41$ & $2014.48 \pm 0.10$ & $52.3 \pm 2.6$ & $0.98616 \pm 0.0014$ & ---\\
 & 191 (S3-223) & 15.1 & $4040 \pm 1100$ & ``\;\;'' & ``\;\;'' & $5543.59 \pm 1100$ & $206 \pm 36$ & $0.99648 \pm 0.00075$ & -180\tn{f}\\
\tabspace
\mr{-3.34} & G2 & --- & $592 \pm 43$ & $1530 \pm 88$ & $1422.68 \pm 43$ & $2014.47 \pm 0.094$ & $57.1 \pm 2.6$ & $0.98811 \pm 0.0010$ & ---\\
 & 230 (S4-23) & 15.9 & $6770 \pm 3400$ & ``\;\;'' & ``\;\;'' & $8220.92 \pm 3400$ & $290 \pm 89$ & $0.99770 \pm 0.00068$ & -100\tn{e}\\
\tabspace
\mr{-4.44} & G2 & --- & $461 \pm 52$ & $2800 \pm 240$ & $1553.19 \pm 52$ & $2014.36 \pm 0.13$ & $47.3 \pm 3.5$ & $0.97523 \pm 0.0021$ & ---\\
 & 95 (S2-198) & 15.6 & $1450 \pm 1300$ & ``\;\;'' & ``\;\;'' & $2991.26 \pm 1300$ & $102 \pm 45$ & $0.98842 \pm 0.0028$ & 40\\
\tabspace
\mr{-6.38} & G2 & --- & $628 \pm 36$ & $1000 \pm 60$ & $1386.08 \pm 36$ & $2014.42 \pm 0.096$ & $59.1 \pm 2.1$ & $0.99260 \pm 0.00050$ & ---\\
 & 126 (S2-84) & 15.3 & $1680 \pm 140$ & ``\;\;'' & ``\;\;'' & $3073.79 \pm 120$ & $114 \pm 5.5$ & $0.99616 \pm 0.00033$ & -60\tn{e}\\
\tabspace
\mr{-6.39} & G2 & --- & $209 \pm 13$ & $865 \pm 170$ & $1805.02 \pm 13$ & $2014.25 \pm 0.13$ & $28.9 \pm 0.98$ & $0.98610 \pm 0.0022$ & ---\\
 & 23 (S1-34) & 13.1 & $836 \pm 250$ & ``\;\;'' & ``\;\;'' & $2641.63 \pm 250$ & $72.7 \pm 12$ & $0.99452 \pm 0.0011$ & -200\tn{f}\\
\tabspace
\mr{-7.42} & G2 & --- & $314 \pm 28$ & $1690 \pm 130$ & $1700.24 \pm 28$ & $2014.31 \pm 0.097$ & $37.4 \pm 2.1$ & $0.97987 \pm 0.0012$ & ---\\
 & 51 (S1-167) & 15.9 & $1180 \pm 770$ & ``\;\;'' & ``\;\;'' & $2888.84 \pm 790$ & $90.5 \pm 31$ & $0.99173 \pm 0.0019$ & -100\\
\tabspace
\mr{-9.45} & G2 & --- & $942 \pm 120$ & $1070 \pm 74$ & $1073.03 \pm 120$ & $2014.48 \pm 0.098$ & $77.7 \pm 6.3$ & $0.99381 \pm 0.00065$ & ---\\
 & 161 (S3-151) & 15.8 & $1750 \pm 87$ & ``\;\;'' & ``\;\;'' & $2821.38 \pm 120$ & $118 \pm 3.7$ & $0.99596 \pm 0.00036$ & -300\tn{f}\\
\tabspace
\mr{-10.6} & G2 & --- & $362 \pm 45$ & $4690 \pm 250$ & $1652.03 \pm 45$ & $2014.23 \pm 0.16$ & $40.3 \pm 3.5$ & $0.95122 \pm 0.0066$ & ---\\
 & 73 (S2-134) & 15.7 & $15200 \pm 13000$ & ``\;\;'' & ``\;\;'' & $16869.2 \pm 13000$ & $485 \pm 250$ & $0.99598 \pm 0.0025$ & -380\tn{f}\\
\tabspace
\mr{-15.5} & G2 & --- & $710 \pm 46$ & $464 \pm 47$ & $1304.02 \pm 46$ & $2014.40 \pm 0.098$ & $64.4 \pm 2.8$ & $0.99680 \pm 0.00029$ & ---\\
 & 246 (S4-67) & 15.8 & $11300 \pm 18000$ & ``\;\;'' & ``\;\;'' & $12666.1 \pm 18000$ & $403 \pm 310$ & $0.99949 \pm 0.00021$ & -200\\
\tabspace
\mr{-22.3} & G2 & --- & $855 \pm 110$ & $1380 \pm 220$ & $1159.77 \pm 110$ & $2014.53 \pm 0.14$ & $72.7 \pm 6.4$ & $0.99172 \pm 0.0012$ & ---\\
 & 144 (S3-20) & 14.6 & $1660 \pm 700$ & ``\;\;'' & ``\;\;'' & $2778.17 \pm 770$ & $112 \pm 26$ & $0.99463 \pm 0.0013$ & 20\tn{e}\\\hline
\multicolumn{10}{c}{\citet{Schodel:2009jm}}\\
\hline
\mr{-7.66} & G2 & --- & $223 \pm 8.6$ & $856 \pm 110$ & $1791.76 \pm 8.6$ & $2014.28 \pm 0.099$ & $30.1 \pm 0.71$ & $0.98688 \pm 0.0016$ & ---\\
 & 45 (S1-34) & 13.2 & $680 \pm 58$ & ``\;\;'' & ``\;\;'' & $2472.16 \pm 58$ & $63.3 \pm 3.6$ & $0.99378 \pm 0.00089$ & -140\tn{f}\\
\tabspace
\mr{-8.25} & G2 & --- & $2230 \pm 85$ & $965 \pm 64$ & $-210.571 \pm 85$ & $2014.57 \pm 0.13$ & $138 \pm 3.6$ & $0.99688 \pm 0.00023$ & ---\\
 & 1136 & 15.5 & $7540 \pm 430$ & ``\;\;'' & ``\;\;'' & $7330.02 \pm 440$ & $312 \pm 12$ & $0.99861 \pm 0.00012$ & -80\\
\tabspace
\mr{-9.49} & G2 & --- & $4800 \pm 220$ & $2220 \pm 140$ & $-2790.29 \pm 220$ & $2014.62 \pm 0.14$ & $228 \pm 6.8$ & $0.99579 \pm 0.00029$ & ---\\
 & 1033 & 15.6 & $6720 \pm 230$ & ``\;\;'' & ``\;\;'' & $3936.34 \pm 98$ & $286 \pm 6.6$ & $0.99664 \pm 0.00024$ & 80\\
\tabspace
\mr{-9.58} & G2 & --- & $7130 \pm 290$ & $1710 \pm 100$ & $-5117.66 \pm 290$ & $2014.62 \pm 0.11$ & $298 \pm 7.9$ & $0.99751 \pm 0.00017$ & ---\\
 & 1258 & 15.8 & $9140 \pm 310$ & ``\;\;'' & ``\;\;'' & $4012.25 \pm 72$ & $352 \pm 7.9$ & $0.99788 \pm 0.00014$ & 80\\
\tabspace
\mr{-10.5} & G2 & --- & $5140 \pm 200$ & $1430 \pm 110$ & $-3125.60 \pm 200$ & $2014.60 \pm 0.12$ & $241 \pm 6.4$ & $0.99736 \pm 0.00021$ & ---\\
 & 2565 & 15.5 & $11700 \pm 440$ & ``\;\;'' & ``\;\;'' & $8536.29 \pm 410$ & $416 \pm 12$ & $0.99847 \pm 0.00014$ & -20\\
\tabspace
\mr{-11.1} & G2 & --- & $2180 \pm 83$ & $1160 \pm 61$ & $-165.054 \pm 83$ & $2014.57 \pm 0.12$ & $137 \pm 3.5$ & $0.99619 \pm 0.00023$ & ---\\
 & 832 & 15.8 & $5190 \pm 200$ & ``\;\;'' & ``\;\;'' & $5024.27 \pm 210$ & $243 \pm 6.4$ & $0.99786 \pm 0.00015$ & -40\\
\tabspace
\mr{-11.4} & G2 & --- & $677 \pm 25$ & $992 \pm 74$ & $1337.07 \pm 25$ & $2014.44 \pm 0.10$ & $62.3 \pm 1.6$ & $0.99299 \pm 0.00053$ & ---\\
 & 154 (S2-84?) & 15.2 & $1540 \pm 77$ & ``\;\;'' & ``\;\;'' & $2875.73 \pm 80$ & $108 \pm 3.5$ & $0.99594 \pm 0.00038$ & -40\\
\tabspace
\mr{-12.3} & G2 & --- & $6840 \pm 280$ & $2750 \pm 120$ & $-4821.53 \pm 280$ & $2014.67 \pm 0.14$ & $288 \pm 8.1$ & $0.99591 \pm 0.00022$ & ---\\
 & 3471 & 14.1 & $14200 \pm 410$ & ``\;\;'' & ``\;\;'' & $9302.72 \pm 370$ & $467 \pm 10$ & $0.99748 \pm 0.00014$ & 0\\
\tabspace
\mr{-12.7} & G2 & --- & $4820 \pm 460$ & $2650 \pm 250$ & $-2805.24 \pm 460$ & $2014.61 \pm 0.14$ & $227 \pm 14$ & $0.99504 \pm 0.00072$ & ---\\
 & 4445 & 15.8 & $54400 \pm 46000$ & ``\;\;'' & ``\;\;'' & $51529.9 \pm 46000$ & $1140 \pm 550$ & $0.99903 \pm 0.00050$ & -120\\
\tabspace
\mr{-13.2} & G2 & --- & $114 \pm 4.6$ & $2510 \pm 170$ & $1900.37 \pm 4.6$ & $2014.06 \pm 0.11$ & $19.1 \pm 0.48$ & $0.94065 \pm 0.0039$ & ---\\
 & 17 (S0-29) & 15.6 & $170 \pm 6.4$ & ``\;\;'' & ``\;\;'' & $2069.83 \pm 4$ & $24.9 \pm 0.61$ & $0.95435 \pm 0.0031$ & 240\\
\tabspace
\mr{-14.9} & G2 & --- & $2910 \pm 110$ & $2610 \pm 130$ & $-897.618 \pm 110$ & $2014.64 \pm 0.13$ & $163 \pm 4$ & $0.99315 \pm 0.00040$ & ---\\
 & 2144 & 15.7 & $15900 \pm 1200$ & ``\;\;'' & ``\;\;'' & $15016.5 \pm 1200$ & $506 \pm 26$ & $0.99779 \pm 0.00017$ & -100\\
\tabspace
\mr{-15.5} & G2 & --- & $17700 \pm 2200$ & $1830 \pm 2000$ & $-15684.2 \pm 2200$ & $2014.68 \pm 1.3$ & $551 \pm 61$ & $0.99851 \pm 0.0030$ & ---\\
 & 5085 & 14.1 & $24800 \pm 25000$ & ``\;\;'' & ``\;\;'' & $9049.01 \pm 27000$ & $690 \pm 200$ & $0.99880 \pm 0.00013$ & 40\\
\tabspace
\mr{-17.7} & G2 & --- & $7800 \pm 250$ & $846 \pm 60$ & $-5784.10 \pm 250$ & $2014.55 \pm 0.14$ & $317 \pm 6.5$ & $0.99882 \pm 0.000088$ & ---\\
 & 5849 & 14 & $57500 \pm 6700$ & ``\;\;'' & ``\;\;'' & $51682.8 \pm 6700$ & $1200 \pm 95$ & $0.99969 \pm 0.000038$ & -60
\end{tabularx}\label{tab:orbits}
\begin{tablenotes}
\small
\item
Note: Each candidate is shown paired with its corresponding set of G2 parameters.
\end{tablenotes}
\tablenotetext{1}{Does not include $v_{z}$, even if available.}
\tablenotetext{2}{Previous encounter date.}
\tablenotetext{3}{Next encounter date.}
\tablenotetext{4}{Model prediction for 2014.0.}
\tablenotetext{5}{Consistent with observed $v_{z}$, S. Gillessen and A. Ghez, priv. comm.; \citet{Paumard:2006fo,Gillessen:2009fg,Do:2013fn}.}
\tablenotetext{6}{Inconsistent with observed $v_{z}$, as above.}
\end{threeparttable}
\end{table*}

Lastly, our fitting includes two additional free parameters to measure any extra variance in G2's position and radial velocity. This is motivated by the change in time in the reported orbital elements of G2, suggesting that the measurement error bars may not capture the full uncertainty in G2's position. This may either arise from deviations from a pure Keplerian orbit, as might be expected when the cloud interacts with the gas surrounding \sa \citep{Abarca:2014ei}, or if the \brg emission does not follow the mass \citep{Phifer:2013bm}. In total, our model includes 22 free parameters.

We then run independent maximum-likelihood analyses (MLAs) using {\tt emcee} \citep{ForemanMackey:2013io} with each MLA presuming that a star of the catalogs of \citet{Schodel:2009jm} or \citet{Yelda:2010ig} may be the star that was tidally disrupted to produce the G2 cloud. Given a typical bolometric correction of $\Sim 2$ in K \citep{Buzzoni:2010df}, the average K extinction in the GC of $\Sim 2.5$ mag \citep{Schodel:2010hx}, and minimum size required for a star to lose mass at G2's periapse $R_{\ast,\min} \sim 0.3$ AU, we exclude stars with ${\rm K} > 16$. We also exclude stars whose proper motions are greater than the escape velocity from \sa at their observed position, and stars with positive declination. In total we consider 1727 (\citeauthor{Schodel:2009jm}) and 512 (\citeauthor{Yelda:2010ig}) stars. For this {\it Letter} our MLAs utilize the positions and velocities of G2 and S2 reported in \citet{Gillessen:2009fg,Gillessen:2013fb,Gillessen:2013bt}\footnote{Publicly available at \url{https://wiki.mpe.mpg.de/gascloud}}, \citet{Ghez:2008hf}, and \citet{Phifer:2013bm}; the combination of these datasets required repeating the alignment procedure of \citet{Gillessen:2009fg} to account for relative proper motion between the two observing frames. We then sort the stars based on their MLA scores.

\subsection{Candidate Late-Type Stars}
Most of the stars of the aforementioned catalogs can be immediately rejected as either their positions or proper motions are not consistent with a star who shares five of six of G2's orbital elements. Some fraction of the stars have orbits that are almost compatible with G2's, but only when allowing $r_{\rm p}$ of the candidate to differ from G2's (i.e. by adding another free parameter). The addition of $r_{\rm p}$ as a free parameter may be physically motivated as G2 could deviate from its original Keplerian trajectory, but observations show that its path is largely consistent with Keplerian and that G2's original periapse does not differ from its measured periapse by more than a factor of $\Sim 2$ \citep{Meyer:2013tb,Phifer:2013bm,Gillessen:2013bt}.

\begin{figure}
\centering\includegraphics[width=\linewidth,clip=true]{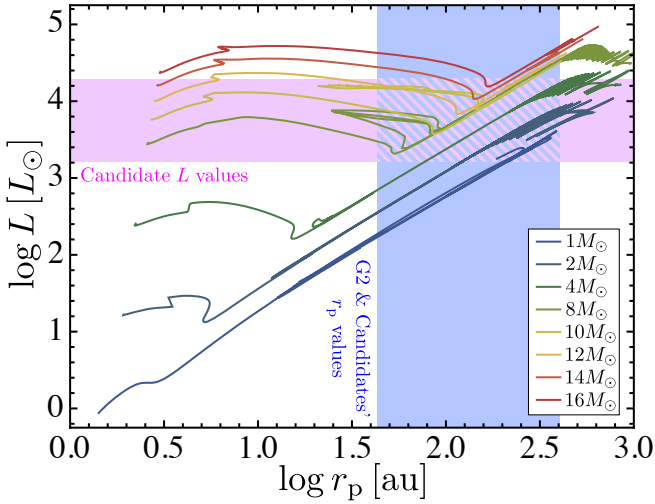}
\caption{Pericenter distances $r_{\rm p}$ of G2 and candidate stars vs. the candidate stars' luminosities $L$. The light blue region corresponds to the range of $r_{\rm p}$ presented in Table \ref{tab:orbits}, whereas the magenta region corresponds to the full range of $L$. Overplotted are evolution tracks that show the $r_{\rm p}$ that giant stars of various masses would lose mass to \sa (i.e. $r_{\rm p} = 2 r_{\rm t}$).}
\label{fig:hr}
\end{figure}

When forcing $r_{\rm p}$ of a candidate to be equal to G2's periapse, we find that several stars are potentially compatible (Table \ref{tab:orbits}). For some of the stars in this list radial velocities are available in \citet{Paumard:2006fo,Gillessen:2009fg,Do:2013fn}, the rest are not in the published literature, although we received $v_{z}$ values for some of the candidates privately from S. Gillessen and A. Ghez. This allows us to definitively eliminate some stars from contention, as noted in the table. Among the list of candidates, S4-23 is the highest-scoring star for which $v_{z}$ is consistent with our model prediction (Figure \ref{fig:orbits}), but several stars (especially those in the \citeauthor{Schodel:2009jm} catalog) do not have known $v_{z}$ values and thus remain viable candidates.

In Figure \ref{fig:hr} we show evolution tracks of various stellar masses generated using {\tt MESA} \citep{Paxton:2011jf,Paxton:2013km} as compared to the distribution of luminosities and median G2 periapse distances for our list of candidates. The intersection of the mass-loss tracks with the allowed hatched region suggests that giants with mass $\lesssim 16 M_{\odot}$ are potentially compatible. By better characterizing the candidate stars it is possible to eliminate candidates on the basis that they could not lose mass at the $r_{\rm p}$ suggested by the MLA, but because a giant star that has lost an appreciable amount of mass can be heated by reaccretion \citep{MacLeod:2013dh}, it is possible that the star is hotter and brighter than the standard sequences.

Recently, it has been noted that the tail-like feature that lies in G2's wake extends far beyond G2's orbital ellipse \citep{Gillessen:2013fb}, which \citeauthor{Meyer:2013tb} argued is evidence that it is unassociated with G2. However, the giant disruption scenario predicts that an extended tail should connect G2 to the star that was disrupted, and this tail should {\it not} be cospatial with G2's trajectory (Figure \ref{fig:stream}), nor possess the same radial velocity. In projection, many of the candidates seemingly lie within a few tenths of an arcsecond from the observed tail. While not proof that the feature is genuinely associated with either G2 or a disrupted star, it is highly suggestive.

\section{Implications}\label{sec:implications}

For stars that appear in both catalogs (e.g., S1-34) there are disagreements in the reported positions and proper motions that lead to different orbital solutions and scores. These disagreements are often larger than the quoted error bars, making it difficult to even definitively associate a star with a listing in both catalogs (e.g., S2-84). The disagreements strongly hint that there may be several stars that scored poorly in our MLAs that may score better with revised positions that better account for systematic uncertainties. Follow-up work should be carefully performed with all available data to better constrain these uncertainties, and thus the viability of potential candidate stars for our giant disruption hypothesis.

The minimum amount of mass the star needs to lose to produce the G2 cloud is approximately equal to the average accretion rate from the disruption of one G2 cloud extended to the time of disruption hundreds of years ago, this suggests that a star would need to lose $\gtrsim 100 M_{\oplus}$ at the time of disruption. This mass is small compared to the amount available in a giant star, suggesting that a giant star could have ``spoon-fed'' \sa for potentially tens of millions of years \citep{MacLeod:2013dh}. This repeated interaction greatly increases the rate G2-like clumps would be detectable, as there are always likely to be a few giant stars on such orbits at any time, and always some clouds forming within their host debris streams.

If the giant disruption scenario for G2 is correct, it suggests that a very large fraction of the material deposited within 100 AU of \sa over the past two centuries may have come from a single star. If this material is capable of circularizing and subsequently accreting onto the black hole, giant disruptions may explain a large fraction of low-level supermassive black hole activity in the local universe.

\acknowledgments 
We thank L. Meyer and S. Gillessen for assistance in interpreting the observational data. This work was supported by Einstein grant PF3-140108 (J.G.), NSF grant AST-1312034 (A.L.), and NSF GFRP (M.M.).

\bibliographystyle{apj}
\bibliography{/Users/james/Dropbox/library}

\end{document}